\documentclass{aa}
\usepackage{graphicx}
\def\approxgt{\mathrel{\hbox{\rlap{\lower.55ex \hbox {$\sim$}}
        \kern-.3em \raise.4ex \hbox{$>$}}}}
\def\approxlt{\mathrel{\hbox{\rlap{\lower.55ex \hbox {$\sim$}}
        \kern-.3em \raise.4ex \hbox{$<$}}}}
\begin{document}
   \title{Is there a hard tail in the Coma Cluster X-ray spectrum?}
 \author{Mariachiara Rossetti
   \inst{1,}\inst{2}
   \and
    Silvano Molendi
   \inst{1}
          }

\institute{
         Istituto di Astrofisica Spaziale e Fisica Cosmica,  via Bassini 15,
         I-20133 Milano, Italy
 \and
         Universit\`a degli Studi di Milano, Dip. di Fisica, via Celoria 16 
         I-20133 Milano, Italy 
         }
\abstract{
We report results from a re-analysis of the BeppoSAX observation of Coma and from the 
analysis of a second, yet unpublished observation of the same object.
From our re-analysis of the first observation we find that the statistical evidence
for a hard tail is about 2$\sigma$. From the analysis of the second observation which, 
thanks to the lower background and the longer exposure time, is characterized by 
a larger signal to noise we find no  evidence for a hard tail. 
From the upper limit on the flux of the hard tail, using the standard Inverse
Compton formulae, we derive a lower limit for the magnetic of  $\sim 0.2-0.4\ \mu\rm{Gauss}$
consistent with Faraday rotation measurements. 
   }
\maketitle
%

\section{Introduction}
It is well known that the intense X-ray emission observed in Galaxy Cluster
is due to a hot ($10^7 $K$< T <  10^8 $K) tenuous 
(n $\sim 10^{-4}$cm$^{-2} -10^{-2}$cm$^{-2}$)  plasma. We also know,
that this plasma is weakly magnetized: Faraday Rotation (FR) Measurements seem to yeld field values of  a few $\mu$Gauss (e.g. Clarke et al., 2001). However 
this last fact has enjoyed relatively little consideration, particularly 
amongst high energy astronomers, barring a few notable exceptions e.g. Goldshmidt \& Rephaeli (1993), as it has little influence on the emission mechanism 
responsible for the bulk of the X-ray emission observed in Galaxy Clusters.\\
While the detection of synchrotron emission at radio frequencies 
unambiguously shows that magnetic fields must be present in clusters, 
the assesment of the intensity of such field is not an easy task. 
In fact, the uncertainty in FR results is great for statistical and physical reasons, due to the randomness in the spatial structure of turbolent magnetic fields (for a more complete description of these problems, see Newman et al., 2002).\\
For the above reasons it would be extremely useful to have another independent
way of measuring B fields in Clusters. The detection of Inverse Compton (IC) emission
of 3K background photons from the relativistic electrons responsible of the 
synchrotron emission observed at radio wavlengths offers, at least  in principle,
an elegant and powerful alternative to asses Cluster magnetic fields, in those objects that host a radio halo.
Up to a few years ago, the search for IC emission in the hard X-ray energy band, the so-called ``hard
tails'', had not met with success (see Rephaeli et al. 1987,  Rephaeli \& Gruber 1998,
Henriksen 1998). Recently  Fusco-Femiano et al. (1999), using BeppoSAX PDS data, and Rephaeli et al. (1999) using RXTE data, find evidences for a non-thermal component in the Coma cluster hard X-ray spectrum. 
Other detections based on BeppoSAX data have been claimed for  
A2256 (Fusco-Femiano et al. 2000) and A754 (Fusco-Femiano et al. 2003).\\
In this Letter we reanalyze the BeppoSAX Coma observation
discussed in Fusco-Femiano et al. (1999), hereafter FF99, 
and analyze a second 
unpublished observation. The outline of the Letter is as follows:
in Sect.\,2 we discuss the data reduction focusing on issues relevant 
to the characterization of weak sources; 
in Sect.\,3 we present our analysis of the two Coma observations;  
finally in Sect.\,4 we summarize our main findings.
Reported errors are 1$\sigma$ unless otherwise stated, 
the adopted cosmological parameters are 
H$_0 = 50$ km s$^{-1}$ Mpc$^{-1}$ and q$_0 = 0.5 $.

\section{PDS Data Analysis}
The PDS instrument uses the rocking collimators technique for background subtraction.  The standard observation strategy is to observe the target with one collimator while the other monitors the background, and to periodically swap them. Two offset positions are available ($3.5^{\circ}$ away from the source), both used for the background subtraction in the standard data processing.  The differential technique used by the PDS instrument allows to detect signals which are much weaker (a few per cent) than the background, a rather uncommon accomplishment for an X-ray experiment. \\
We have processed data using the standard procedure of the \textsc{Saxdas} package. This procedure does not deal with two possible problems that may have a strong impact on the spectra of weak sources: the presence of an instrumental residual not completely removed by the background subtraction procedure and the possible contamination from sources in the OFF fields. In Sects.\,2.1 and 2.2 we describe the procedure we used to deal with these problems. 

\subsection{Instrumental background residual}
Since clusters of galaxies are extremely weak sources in hard X-rays, it is of great importance to be very careful in the background subtraction. We have checked for the robustness of the standard procedure, analyzing the spectra of 15 ``blank fields'', i.e. fields which do not contain sources showing significant emission in the PDS energy range.
 We have summed spectra from these observations, finding that, although these fields do not contain any individually detected sources, the spectrum differs from zero: the count rate is $(1.45 \pm 0.77)\times 10^{-2} \rm{cts\, s^{-1}}$ in the 12-100 keV energy range. This indicates that the background in the ON position is somewhat larger (at a $2\sigma$ confidence level) than that in the OFF position. In this way we have produced the spectrum of the instrumental contribution which is not removed by the background subtraction procedure. Comparing this spectrum with those of the two observations of the Coma cluster, that we will describe in Sect.\,3, we see that the level of the instrumental residual is comparable to that of the source spectrum, for energies greater than 50 keV. Therefore, it is useful to subtract the instrumental residual from the source spectrum, in order to minimize the instrumental contribution.\\

\subsection{Contamination from sources in the OFF fields}
In order to detect possible sources in the background fields of the Coma observations, we have first evaluated the systematic differences between the two OFF fields which are used for background subtraction. Our analysis, based on 69 fields (see Appendix), has shown that there is a significant difference, of instrumental origin, between the count rate measured in the OFF fields; the same fact has recently been observed indipendentely by Nevalainen et al. (2003). 
 The mean value of the difference between the count rates in the background positions is $\langle\rm{OFF}^- - \rm{OFF}^+\rangle=(2.44 \pm 0.33)\times10^{-2}$ cts/s in the 25-80 keV energy range, while the dispersion around the mean is $\sigma=4.33 \times10^{-2}$ cts/s. A significant contribution to the width of this distribution is given by fluctuations in the Cosmic X-Ray background. \\
This analysis can be a useful tool to detect sources in the OFF fields of observations we want to study: comparing the count rate of the difference between the OFF fields with the distribution of our sample, we can discriminate between statistical fluctuations and contaminating sources.\\
We have applied this procedure to both observations of the Coma cluster: for the first observation we find $\rm{OFF}^- - \rm{OFF}^+=(2.98 \pm 3.65)\times10^{-2}$ cts/s in the 25-80 keV energy range, close to the mean value of our sample. For the second observation, we find a lower value  $\rm{OFF}^- - \rm{OFF}^+=(-2.34 \pm 1.85)\times10^{-2}$ cts/s, which is however consistent with being a fluctuation, since the difference with the mean value is only 1.1 $\sigma$.  

\section{Results}
The Coma Cluster of Galaxies was observed twice by BeppoSAX (see Table 1 for observation log).
\begin{table}[!htbp]
\caption{Observation log}
\footnotesize
\begin{center}
\begin{tabular}{c c c}
\hline
\hline
 & Date & Exposure time\\
\hline
Observation 1$^a$ & 28/12/1997 & 84 ksec\\
Observation 2 & 31/12/2000 & 250 ksec\\
\hline
\end{tabular}
\end{center}
Notes:  $^a$Fusco-Femiano et al. (1999)
\end{table}
In Table 2 we report results for both observations and present best fits with and without  the subtraction of the instrumental residual (Sect.\,2.1).  
Since the instrumental residual may give a significant contribution in spectra at energies greater than 50 keV, we discuss results after the subtraction in more detail.

\subsection{First observation}

Fitting the data with a bremsstrahlung model, we find $kT=(9.17 \pm 0.58)$ keV, which is consistent with recent XMM results (8 keV $< kT <$ 9 keV, Arnaud et al., 2001). As in FF99, we have fixed $kT=8.21$ keV, which is the value obtained by GINGA (Hughes et al., 1993), because this satellite has a field of view which is similar to that of the PDS. We find an excess above the thermal model (see Fig.\,1 and Table 2) at the 2.5 $\sigma$ level and the flux due to the non-thermal component is $(10.46 \pm 4.63)\times 10^{-12} \rm{erg\,cm^{-2} s^{-1}}$ (20-80 keV). This excess is less significant than previously reported ($4.5\,\sigma$ by FF99) and the flux is significantly lower. These differences are due to an error in the data processing on which the previous paper is based. After removing this error, the spectrum extracted with the \textsc{Xas} software is consistent with our spectrum (M.Orlandini, private communication).\\
   \begin{figure}
   \centering
\includegraphics[angle=270,width=8.0cm]{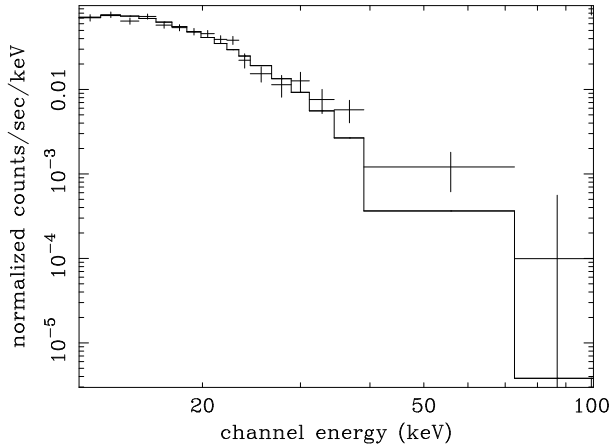}
      \caption{ Spectrum from observation 1 fitted with a bremsstrahlung model with $kT=8.21$ keV.
         }
   \end{figure}
%
The choice of fixing the temperature to the GINGA result ($kT=8.21$ keV) in our fit is somewhat arbitrary, since different values of this parameter have been reported by many instrument ($7.5\pm 0.2$ keV by TENMA, Hughes et al. 1988b, $8.5 \pm 0.3$ keV by EXOSAT, Hughes et al., 1988a, $8.21 \pm 0.16$ keV by GINGA, Hughes et al., 1993). Imaging telescopes revealed large-scale inhomogenities of the temperature in the range 5--11 keV (ASCA, Honda et al., 1996) and more recently have reported the temperature profile of the Coma Cluster, with values in the range 8.5-10 keV (BeppoSAX, De Grandi \& Molendi, 2002) or 8-9 keV (XMM, Arnaud et al. 2001). 
If, on the basis of the above discussion, we leave $kT$ free to vary between 8 keV and 9 keV,  the significance of the excess drops to 2.09 $\sigma$.

\subsection{Second observation}
We have analyzed the second observation of the Coma Cluster performed by BeppoSAX, which has not been published so far. This observation is longer (250 ksec)  than the first one and has a lower background level, because SAX's orbit has lowered with time. For these reasons the signal to noise ratio in the second observation is better than in the first one. \\
   \begin{figure}
   \centering
\includegraphics[angle=270,width=8.0cm]{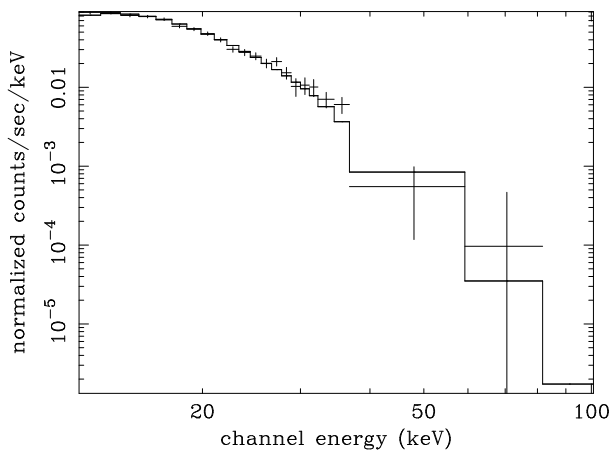}
      \caption{ Spectrum from observation 2 fitted with a bremsstrahlung model  with $kT=8.21$ keV. There is no evidence of an excess above the thermal model at high energies
         }
   \end{figure}
Fitting the data with a thermal bresstrahlung model we find $kT=(8.150\pm 0.20)$ keV, which is consistent with the GINGA result, that we used for the first observation. Fixing $kT=8.21$ keV (Fig.\,2), the excess is significant at less than 1 $\sigma$ level. In Sect.\,2.2, we have noted that in this observation the count rate in the OFF$^+$ position is slightly larger than in the OFF$^-$. Even if this difference is probably a statistical fluctuation, we have analyzed the spectra obtained using only one of the OFF fields as background: results do not vary significantly. In fact, using only the OFF$^-$ field as background, we find $kT=(8.50 \pm 0.27)$ keV and an excess at 1.7 $\sigma$ level above the thermal emission (if we fix the temperature at 8.21 keV), while if we use only the OFF$^+$ field,  $kT=(8.15 \pm 0.21)$ keV and the excess is significant at 0.5 $\sigma$. These values are consistent with those obtained using both OFF fields (see Table 2).\\
 In this observation, there is no need to add a non thermal component and we can only give an upper limit to the flux due to IC scattering. The upper limit to the IC flux depends strongly on the value we assume for the temperature of the cluster: if we assume 8 keV$<kT<$9 keV, the upper limit to the IC flux in the 20-80 keV energy band at the 99\% confidence level, is in the range $3.2\times10^{-12}-8.1\times10^{-12}\ \rm{erg\ cm^{-2}s^{-1}}$.\\
We can use this result to give a lower limit to the magnetic field of the ICM. Comparing the upper limit to the IC flux with the synchrotron flux from the Coma radio-halo we can calculate the lower limit for the magnetic field. Using the expression in Sarazin (1988), we find $B_{\rm{min}} \simeq 0.23-0.37\, \mu\rm{Gauss}$.
This value is consistent with results obtained with FR measurements (Feretti et\ al., 1995), that suggest the presence of two components in the Coma magnetic field: a strong component ($\sim 6\ \mu\rm{Gauss}$) tangled on small scale (less than 1 kpc) and a weak one  ($\sim 0.2-0.3\ \mu\rm{Gauss}$)  on large scale ($\sim 200$ kpc).\\
We have also analyzed the sum of the two observations: results from this analysis are consistent with those of the second observation. However, because of the better statistics that we have in the second observation thanks to the lower background level, with the second observation alone we can give a tighter upper limit to the IC flux, and therefore a tighter lower limit for the magnetic field.\\
We have compared results from this observations with those obtained with a second RXTE observation by Rephaeli \& Gruber (2002).
If we fit our data with their model (thermal model with $kT=7.67$ keV and a power law with $\Gamma =2.1$), the fit is acceptable; the difference in the estimates of the non-thermal flux are mainly due to the different values of the temperature we use in the fit.

\begin{table}[!htbp]
\caption{Parameters of the fits for both observations with bremsstrahlung model}
\footnotesize
\begin{center}
\begin{tabular}{c c c c c}
\hline
\hline
 & $kT$ & excess$^a$ & $\chi^2$ &  d.o.f.$^b$  \\ 
 & (keV) & ($\sigma$) & & \\
\hline
\multicolumn{5}{c}{First observation}\\
\hline
No Subtraction & $9.12 \pm 0.52$ & & 80.7 & 76 \\
After Subtraction & $9.17 \pm 0.58$  & & 78.7 & 76\\
\hline
No Subtraction & 8.21 & 2.84  & 83.8 & 77 \\ 
After Subtraction & 8.21 & 2.51  & 81.6 & 77\\
\hline
\multicolumn{5}{c}{Second observation}\\
\hline
No Subtraction & $8.31 \pm 0.22$ & & 103.0 & 76 \\
After Subtraction & $8.15 \pm 0.20$ & & 91.1 & 76 \\
\hline
No Subtraction & 8.21 & 1.11  & 103.2 & 77 \\ 
After Subtraction & 8.21  & 0.82   & 91.2 & 77 \\
\hline
\end{tabular}
\end{center}
Notes:  $^a$ Estimated as the difference between the observed count rate and the model predicted one in the 25-80 keV energy range, in units of the error on the observed count rate. $^b$ degrees of freedom.
\end{table}

\section{Summary}
In this Letter we have reanalyzed the first PDS Coma observation 
and performed the analysis of a second unpublished Coma observation.
To better asses the presence or absence of a hard tail we have 
performed a detailed study of some issue concerning the PDS background.
More specfically we have assesed the systematics involved in the standard 
background subtraction procedure applied to PDS data and devised 
a technique to acertain the presence of contaminating sources in the 
PDS background fields. Since both of these issues are of interset
to anyone analyzing weak source with the BeppoSAX PDS data we have made 
the necessary files and documentation available at the 
WEB site: \verb|http://www.mi.iasf.cnr.it/~rossetti/PDS_back|\verb . \\
Our reanalysis of the Coma observation published in FF99 shows 
a modest evidence for a hard tail  if the temperature is fixed at the 
value measured with  GINGA, 8.21 keV. 
The significance published in FF99
is inconsistent with the one reported here: the difference is due to an 
error in the reduction of the PDS data on which the  FF99
paper is based. 
We have found that, thanks to its longer exposure time and lower background intensity, the 
second unpublished Coma observation is better suited to investigate the presence of a hard tail.
Our analysis shows that the observed spectrum is consistent with a thermal model with a 
temperature of $(8.15 \pm 0.25)$ keV in agreement with the GINGA measurement.
Allowing the temperature to vary between 8 keV and 9 keV we have constrained the  upper limit 
on the hard tail in the range $3.2 \times 10^{-12} - 8.1  \times 10^{-12}$ erg cm$^{-2}$ s$^{-1}$,
using the standard Inverse Compton argument we have converted these measurement into
a lower limit on the B field of $\simeq 0.23-0.37\ \mu$Gauss.
This result is consistent with FR measurements (Feretti et al. 1995) which indicate
that the B field in Coma is structured in a strong component ($\simeq 6\ \mu\rm{Gauss}$) ordered
on small scales (less than 1 kpc) and a weak one  ($\simeq 0.2-0.3\ \mu\rm{Gauss}$)  on large 
scales ($\sim 200$ kpc).
Future prospects for the detection of the IC emission from Coma depend very much upon 
which of the above two components is responsible for the synchrotron emission observed in
this cluster. If the emission is associated to the weak component future experiments 
with a sensitivity only a factor of a few better than that of the BeppoSAX PDS will
detect the IC emission. If, on the contrary, the emission is associated to the more intense 
B field component, then an instrument with a sensitivity $10^{4}$ times better than the PDS
is required. Such an experiment is unlikely to be flown in the near and not so near future.

\section*{Appendix: Systematic differences between tha background fields}
In order to evaluate the  systematic  differences between tha background fields we have analyzed a sample of  observations performed by the PDS instrument. Our sample consists of 69 observations whose  target is outside the galactic plane ($|b|>27^{\circ}$) and with a long exposure time.\\
 We have subtracted the count rate observed in the OFF$^-$ position from that of the OFF$^+$ position for each of the 4 detectors of the PDS and we find that the mean values significantly differ from zero and depend strongly on collimators. In fact, the mean values $\langle\rm{OFF}^- - \rm{OFF}^+\rangle$ are positive for detectors 1 and 2, associated with collimator A, and negative for detectors 3 and 4 (collimator B). Also the mean value  $\langle\rm{OFF}^- - \rm{OFF}^+\rangle$ for the whole instrument is significantly different from zero and positive ($\langle\rm{OFF}^- - \rm{OFF}^+\rangle=(2.44 \pm 0.33)\times10^{-2}$ cts/s in the 25-80 keV energy range) while the dispersion around the mean is $\sigma=4.33 \times10^{-2}$ cts/s.

\begin{acknowledgements}
We thank the referee  Yoel Rephaeli for a critical reading of the manuscript.
\end{acknowledgements}

{}

\end{document}